%% file: paper.tex
\documentclass[preprint]{aastex}
\input epsf.sty

\slugcomment{Submitted to ApJ}

\newcommand\al{\alpha}
\newcommand\bt{\beta}
\renewcommand\th{\theta}
\newcommand\m{M}

\newcommand\phiij{\phi_{ij}}
\newcommand\phixx{\phi_{xx}}
\newcommand\phiyy{\phi_{yy}}
\newcommand\phixy{\phi_{xy}}

\newcommand\Hunits{\mbox{km s$^{-1}$ Mpc$^{-1}$}}
\newcommand\gravlens{{\it gravlens\/}}
\newcommand\lensmodel{{\it lensmodel\/}}

\newcommand\xx{{\bf x}}
\newcommand\xpp{{\bf x}'}
\newcommand\yy{{\bf y}}
\newcommand\uu{{\bf u}}
\newcommand\bb{{\bf b}}
\newcommand\pp{{\bf p}}
\newcommand\qq{{\bf q}}

\newcommand\del{{\bf\nabla}}

\newcommand\refeq[1]{eq.~(\ref{eq:#1})}
\newcommand\refEq[1]{Eq.~(\ref{eq:#1})}

\newcommand\reffig[1]{Figure~\ref{fig:#1}}
\newcommand\reffigs[2]{Figures~\ref{fig:#1} and \ref{fig:#2}}
\newcommand\reftab[1]{Table~\ref{tab:#1}}

\begin{document}

\title{Computational Methods for Gravitational Lensing}
\author{Charles R.\ Keeton}
\affil{Steward Observatory, University of Arizona, \\
933 N.\ Cherry Ave., Tucson, AZ 85721}

\begin{abstract}
Modern applications of strong gravitational lensing require the
ability to use precise and varied observational data to constrain
complex lens models. I discuss two sets of computational methods
for lensing calculations. The first is a new algorithm for solving
the lens equation for general mass distributions. This algorithm
makes it possible to apply arbitrarily complicated models to
observed lenses. The second is an evaluation of techniques for
using observational data including positions, fluxes, and time
delays of point-like images, as well as maps of extended images,
to constrain models of strong lenses. The techniques presented
here are implemented in a flexible and user-friendly software
package called \gravlens, which is made available to the community.
\end{abstract}

\section{Introduction}

Gravitational lensing is an important astrophysical tool because it
directly probes mass (as opposed to luminosity) distributions, and
because it brightens and enlarges the images of distant sources.
The more than 60 known strong lenses\footnote{Strong lenses are
systems with multiple images of a background source, and they are
the focus of this paper. Weak lensing, or shape distortions without
multiple imaging, also probes mass distributions but with different
techniques. For recent reviews of weak lensing, see Mellier (1999)
and Bartelmann \& Schneider (2001).} produced by galaxies
constitute a {\it mass-selected\/} sample of galaxies at
intermediate redshifts. Lensing provides precise mass measurements
for these galaxies, and thereby offers a powerful probe of the
physical properties of intermediate-redshift galaxies and the
evolution of early-type galaxies in low-density environments (e.g.,
Keeton, Kochanek \& Falco 1998; Kochanek et al.\ 2000). The lenses
can be used to probe galaxy mass distributions (e.g., Kochanek
1991a; Rusin \& Tegmark 2000; Rusin \& Ma 2000; and references
therein). They can also be used for direct measurements of the
Hubble constant independent of the distance ladder (e.g., Koopmans
\& Fassnacht 1999; Witt, Mao \& Keeton 2000; and references
therein), to constrain the cosmological model (e.g., Falco,
Kochanek \& Mu\~noz 1998; Helbig et al.\ 1999; and references
therein), and for detailed studies of the host galaxies of
high-redshift quasars (e.g., Rix et al.\ 2000; Kochanek, Keeton \&
McLeod 2001a). Strong lenses produced by clusters probe the radial
mass distribution of clusters and reveal the clumpy galaxy
distribution superimposed on the smooth cluster background, and
thereby test models of structure formation in the cold dark matter
paradigm (e.g., Tyson, Kochanski \& Dell'Antonio 1998; Williams,
Navarro \& Bartelmann 1999; Shapiro \& Iliev 2000).

In many of these eclectic lensing applications an essential step is
fitting mass models to observed lenses. There are two key
ingredients to modern lens models. The first is the ability to use
the precise and varied observational data now available for many
lenses. Optical and near-infrared astrometry with the Hubble Space
Telescope achieves a precision of a few milli-arcseconds (e.g.,
Leh\'ar et al.\ 2000).  Radio astrometry from VLBI or VLBA maps
can achieve a precision of 10 micro-arcseconds or better, and may
resolve fine substructure in the images (e.g., Patnaik, Porcas \&
Browne 1995; Trotter, Winn \& Hewitt 2000). Deep optical and
especially near-infrared images often reveal extended structure
due to the host galaxy of the source, and even some complete
Einstein rings (e.g., Bernstein et al.\ 1997; Impey et al.\ 1998;
Kochanek et al.\ 2001a). Photometry at many wavelengths and many
epochs can reveal evidence for reddening and/or microlensing of
the images (e.g., Gott et al.\ 1981; Falco et al.\ 1999;
Wambsganss et al.\ 2000; Wozniak et al.\ 2000), and the monitoring
can be used to measure time delays between the lensed images (e.g.,
Kundi\'c et al.\ 1997a; Schechter et al. 1997). With the proper
techniques, all of the different kinds of data can be used to
constrain lens models.

The second important ingredient is the ability to study complex
mass models. Reproducing qualitative features of a lens (the number
and configuration of the images) can often be done with very simple
models, but fitting the data quantitatively requires models that
include detailed structure in the lens galaxy and its environment.
For example, the models require an elliptical density distribution
for early-type lens galaxies (e.g., Keeton \& Kochanek 1997), or a
thin disk and a rounder halo for spiral lens galaxies (e.g.,
Maller, Flores \& Primack 1997), and perhaps even substructure in
the galaxy (e.g., Mao \& Schneider 1998; Bernstein \& Fischer
1999). The models usually must include tidal perturbations from
objects near the lens galaxy or along the line of sight (e.g.,
Young et al.\ 1981; Keeton, Kochanek \& Seljak 1997; Witt \& Mao
1997). While tidal perturbations are often approximated as an
external shear for convenience, this simplification may be ruled
out by data of sufficient quality (e.g., Impey et al.\ 1998).
Several lenses are even found to have multiple galaxies that lie
inside the Einstein ring and must be explicitly modeled (e.g.,
Koopmans \& Fassnacht 1999; Rusin et al.\ 2000). The problem is
that such complicated models can be hard to study due to the
difficulty of solving the lens equation. For models with spherical
or ellipsoidal symmetry all classes of solutions to the lens
equation are known (see, e.g., Schneider, Ehlers \& Falco 1992).
For models without such symmetry, however, it may not even be
clear what the maximum number of images is, much less how they
are arranged. The need to use complex models to fit the data
means that we need a general algorithm for solving the lens
equation without requiring simplifying assumptions about symmetry.

This paper presents methods for modern lensing calculations,
including a fully general algorithm for solving the lens equation
that requires no assumptions about the symmetry of the mass models,
and techniques for handling a variety of observational data. The
organization of the paper is as follows. Section 2 reviews the lens
theory needed for most calculations. Section 3 presents the
algorithm for solving the lens equation, while Section 4 discusses
techniques for using observational data to constrain lens models.
The methods presented here are implemented in a publicly-available
software package called \gravlens, which is described in Section 5.
The software and documentation are available from the web site of
the CfA/Arizona Space Telescope Lens Survey, at
{\tt http://cfa-www.harvard.edu/castles}.

\section{Basic Lens Theory}

Suppose a source at angular position $\uu$ emits a light ray that
passes a foreground mass distribution (the lens) with impact
parameter $\xx$ and gets deflected by the gravitational field of
the lens. Compared to an undeflected ray, the deflected ray has a
longer travel time because it has a longer geometric length and it
passes through a gravitational potential well. Assuming that the
lens is confined to a small fraction of the total path length, the
extra light travel time is
\begin{equation} \label{eq:tdel}
  \tau(\xx) = {1+z_l \over c}\,{D_{ol} D_{os} \over D_{ls}}
  \left[{1\over2}\left|\xx-\uu\right|^2 - \phi(\xx)\right] ,
\end{equation}
where $z_l$ is the redshift of the lens, and $D_{ol}$, $D_{os}$,
and $D_{ls}$ are angular diameter distances from the observer to
the lens, from the observer to the source, and from the lens to
the source, respectively. (See, e.g., Schneider et al.\ 1992 for
a full discussion.) Also, $\phi$ is the two-dimensional
gravitational potential of the lens,
\begin{equation} \label{eq:phi}
  \phi(\xx) = {1\over\pi}\,\int
    { \Sigma(\yy) \over \Sigma_{cr} }\,
    \ln|\xx-\yy|\, d\yy\,,
\end{equation}
where $\Sigma$ is the surface mass density of the lens, which is
normalized by the critical surface density for lensing,
\begin{equation} \label{eq:Sigcr}
  \Sigma_{cr} = {c^2 \over 4\pi G}\,{D_{os} \over D_{ol} D_{ls}}\ .
\end{equation}
By Fermat's principle, images form at stationary points of the time
delay surface, or at solutions of the equation
\begin{equation} \label{eq:lens}
  \uu = \xx - \del\phi(\xx)\, .
\end{equation}
This is the general form of the gravitational lens equation (for a
single lens plane).

In addition to producing a deflection, the lens also distorts and
amplifies the image(s) in a manner described by the magnification
tensor,
\begin{equation} \label{eq:mag}
  \mu \equiv \left({\partial\uu \over \partial\xx}\right)^{-1}
  = \left[\begin{array}{cc}
    1 - \phixx & -\phixy \\
    -\phixy & 1 - \phiyy \\
  \end{array}\right]^{-1} ,
\end{equation}
where subscripts denote partial differentiation, $\phiij \equiv
\partial^2 \phi / \partial x_i \partial x_j$. Generically, a strong
lens has one or more ``critical curves'' in the image plane along
which $\det(\mu^{-1}) = 0$. The critical curves map to ``caustics''
in the source plane, which are important because they are places
where the number of images changes, and because a source near a
caustic is highly amplified and distorted.

\section{Solving the Lens Equation}

Reading the lens equation from left to right --- i.e., picking
a source $\uu$ and trying to find the images $\xx_i$ --- makes
it hard to solve. Analytic solutions are often impossible because
the deflection $\del\phi$ either involves transcendental functions
or cannot be computed analytically. Numerical solutions are often
difficult because there is no algorithm that is guaranteed to find
all the roots of a two-dimensional equation (see Press et al.\
1992).  Hence, numerical techniques require independent knowledge
of the number of images and rough guesses for their positions,
but no local property of the lens equation gives this information.
While the global caustic structure does characterize the number
of images, it is hard to analyze except in relatively simple lens
models with sufficient symmetry. The existence of compound lenses
(e.g., Koopmans \& Fassnacht 1999; Rusin et al.\ 2000) and lens
galaxies with companions or satellites (e.g., Hogg \& Blandford
1994; Kundi\'c et al.\ 1997b; Tonry 1998) indicates the need for a
general solver that does not require symmetry assumptions.

\begin{figure}
\centerline{\epsfxsize=5.0in \epsfbox{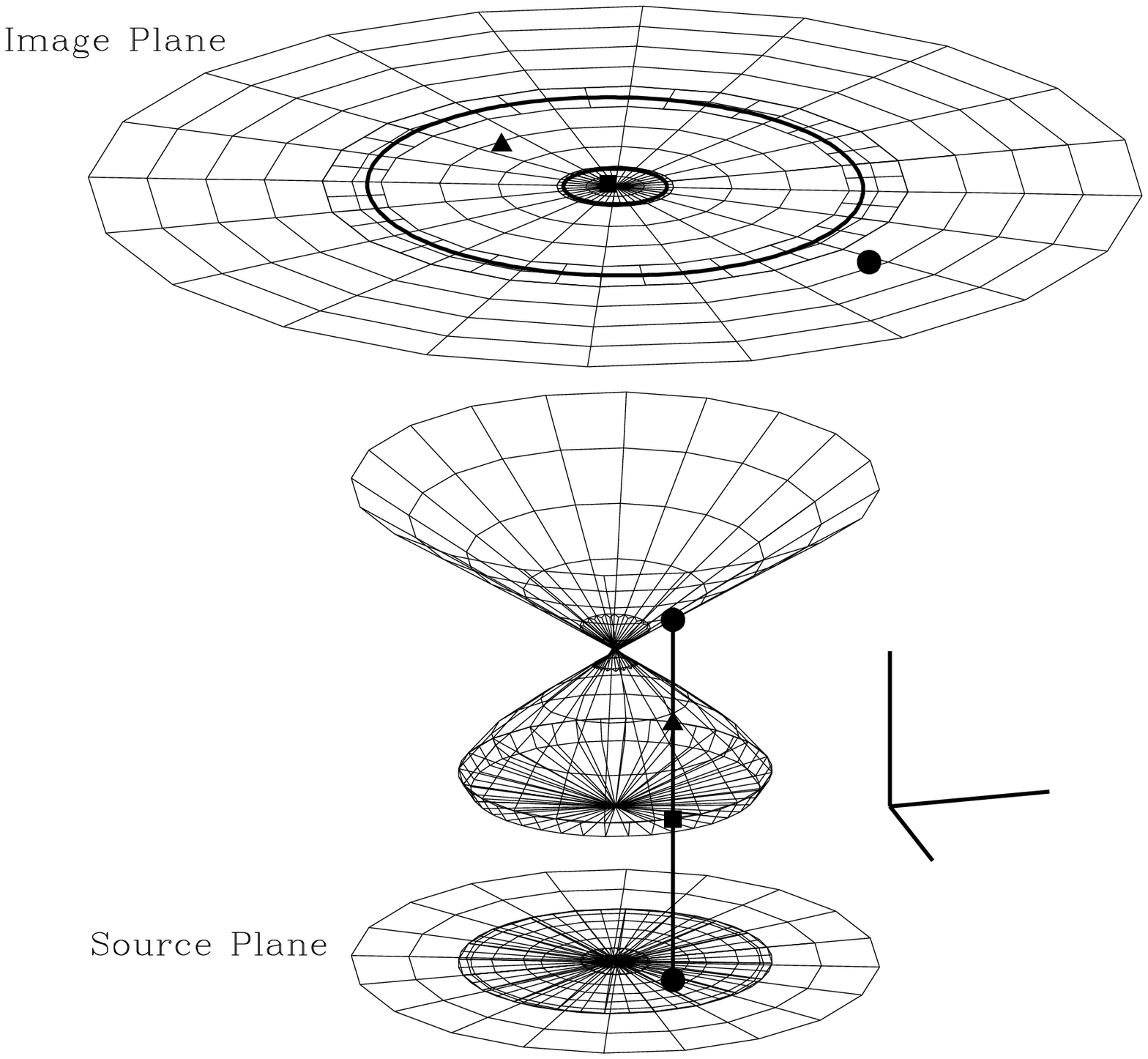}}
\caption{
The geometry of the lens mapping for a nearly circular lens model.
The top figure shows the tiling of the image plane. The middle
figure shows how the image plane is distorted by the lens mapping.
Height has been added to make the folding apparent. Projecting down
through the distorted surface gives the tiling of the source plane
(bottom), which overlaps itself to form multiply-imaged regions.
The heavy curves in the image plane show the critical curves. A
sample source is shown, together with the points where it
intersects the distorted surface and where those points end up in
the image plane. The central image is hard to see because the
tiling is dense.
}\label{fig:cmap}
\end{figure}

The key simplification is to reverse direction and read the lens
equation from right to left, to view it as a mapping from the
image plane to the source plane: it takes each image position
$\xx$ and maps it to a unique source position
$\uu(\xx) = \xx - \del\phi(\xx)$. This way of thinking leads to
a straightforward solution to the problem that a numerical root
finder needs to know the number and approximate locations of the
images. Consider laying down some tiling of the image plane. The
lens mapping takes each image plane tile $I_j$ to a corresponding
source plane tile $S_j$ (by mapping vertices), and thus generates
a tiling of the source plane that covers every point with at least
one tile. \reffig{cmap} depicts a sample tiling. The tiling actually
contains all of the information we need to solve the lens equation.
There is some set of tiles $(S_j,S_k,\ldots)$ that cover any
particular source position, and the number of covering tiles gives
the number of images of that source. Furthermore, the corresponding
image plane tiles $(I_j,I_k,\ldots)$ bound the image
positions.\footnote{If a source is covered by source tile $S_j$,
then an image must lie within the corresponding image plane tile
$I_j$.} In other words, the tiling easily reveals both the number
of images and their approximate positions, which can be efficiently
refined with a numerical root finder.

\begin{figure}
\centerline{\epsfxsize=6.0in \epsfbox{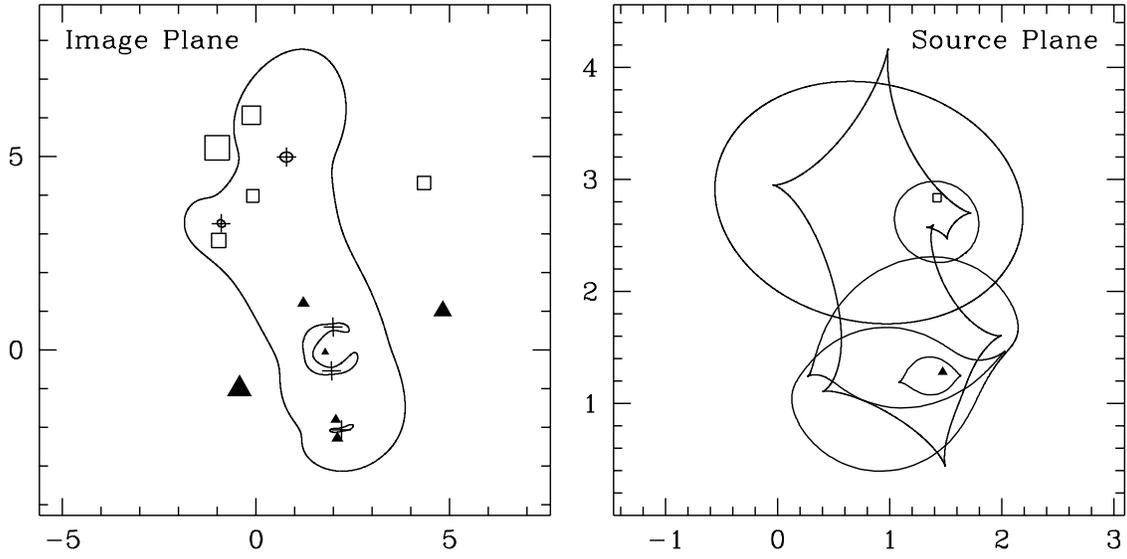}}
\caption{
Sample critical curves and image configurations for a set of five
random galaxies, computed with the tiling algorithm. In the image
plane, the curves show the critical curves, the crosses mark the
five galaxies, and the squares and triangles denote sample images.
In the source plane, the curves show the caustics, and the square
and triangle indicate the two sample sources. The square has seven
images, of which five are bright and two are trapped in galaxy
cores. The triangle has nine images, of which six are bright and
three are trapped in galaxy cores. In the image plane, the size of
each point is proportional to the magnification of the image. The
axis scale is nominally arcseconds but is in fact arbitrary, because
this example is intended only to demonstrate the capabilities of
the tiling algorithm.
}\label{fig:random}
\end{figure}

The lens mapping incorporates all of the properties of the lens
equation, such as the folding and stretching that determine the
number of images and their distortions (see \reffig{cmap}). The
tiling automatically includes all of the features of the lens
mapping (albeit at finite resolution), without requiring any
assumptions about symmetry in the lensing mass. As a result, the
tiling algorithm provides a fully general method for solving the
lens equation to analyze the lensing properties of arbitrary mass
distributions. Blandford \& Kochanek (1987) and Kochanek \&
Blandford (1987) have used a similar technique for calculations
of lens statistics. More generally, the tiling algorithm can be
used to find the critical curves and caustics and to solve the
lens equation for arbitrarily complicated mass distributions, such
as the random collection of five galaxies shown in \reffig{random}.

Implementing the tiling algorithm reveals three important details.
First, the tiles should be triangles, because triangles are the
only polygons that are guaranteed to remain convex regardless of
how they are deformed by the mapping. Also, triangular tiles make
it easy to determine whether a given source point $\uu$ is covered
by a tile. Let $\uu_i$ $(i=1,2,3)$ be the vertices of the tile, and
let $\delta\uu_i = \uu_i - \uu$ be the vector from the source point
to vertex $i$; the source is contained within the triangle if the
three cross products $\delta\uu_1 \times \delta\uu_2$, $\delta\uu_2
\times \delta\uu_3$, and $\delta\uu_3 \times \delta\uu_1$ all have
the same sign. In practice, it may be convenient to use
quadrilateral tiles (see \reffig{cmap}) and divide each one into
two triangles.

\begin{figure}
\centerline{\epsfxsize=6.0in \epsfbox{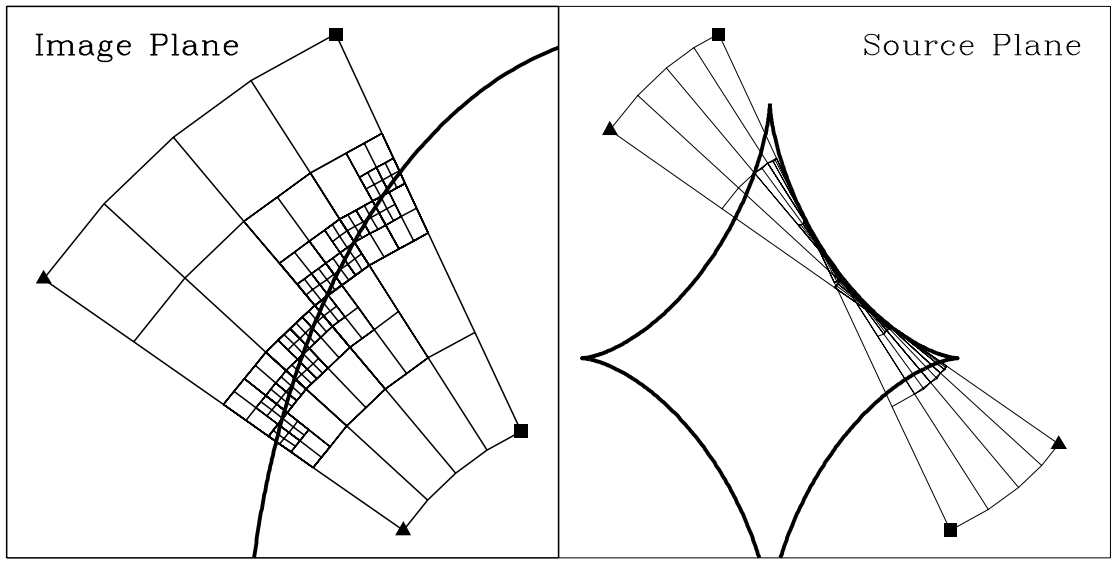}}
\caption{
A sample tiling near a fold caustic, with three levels of
sub-tiling. The left panel shows part of the image plane, and the
right panel shows the corresponding part of the source plane. The
heavy curves show the critical curve and caustic. The filled
triangles and squares help illustrate the fold by matching vertices
in the image and source planes to each other.
}\label{fig:subt}
\end{figure}

Second, we must consider the resolution of the tiling. Good
resolution is important near the critical curves in order to
resolve the folding (see \reffigs{cmap}{subt}), but to avoid
unnecessary calculations we want an adaptive algorithm that uses
high resolution only where necessary. Fortunately there is a simple
local criterion that identifies such regions. If the scalar
magnification, $\det(\mu)$, changes sign across an image plane tile
then that tile contains a critical curve. To increase the
resolution, the tile can be broken into an array of sub-tiles. This
sub-tiling can then be recursively repeated to obtain even better
resolution. \reffig{subt} illustrates a tiling that has three
levels of sub-tiling near the critical curves. In lens models with
more than one galaxy, recursive sub-tiling can also be used near
the additional galaxies to resolve their critical curves. The
sub-tiling offers the additional benefit of putting tight bounds
on the locations of the critical curves, which can then be refined
with a numerical root finder.

\begin{figure}
\centerline{\epsfxsize=4.5in \epsfbox{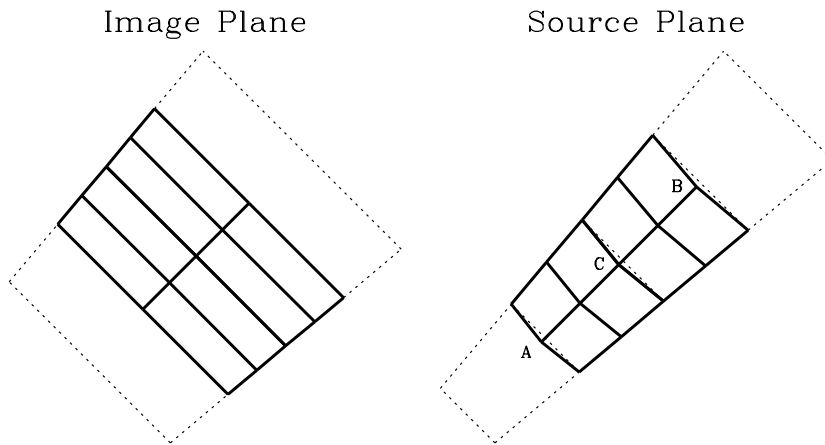}}
\caption{
A sample tiling with four tiles, two of which have $2\times2$
sub-tiling; the light dotted lines show the main tiling, and the
heavy solid lines show the sub-tiling. In the image plane the
vertices of a daughter tile lie on the edges of its parent tile,
but this may not be true in the source plane.  The result can
be an overlap (point A) or a gap (point B) in the source plane
tiling. A gap or overlap appears only at places where the tiling
changes resolution; point C shows that if adjacent tiles both
have sub-tiles, the gap on one side is exactly offset by the
overlap on the other.
}\label{fig:gaps}
\end{figure}

Finally, the sub-tiling introduces one additional technicality.
Consider placing the vertex of a daughter tile on the edge of its
parent tile. Because a straight line in the image plane may map to
a curve in the source plane, in the source plane the daughter
vertex map {\it not\/} lie on the edge of the parent tile. As a
result, the daughter tiles may not cover exactly the same area as
the parent tile; the result can be gaps or overlaps in the tiling,
as illustrated in \reffig{gaps}. This problem occurs only at places
where the resolution changes. (If adjacent tiles are both
sub-tiled, the gap in one sub-tiling is exactly offset by the
overlap in the other; see point C in \reffig{gaps}.) The problem
can be easily corrected by using $2\times2$ sub-tiling so that each
gap or overlap is a triangle that can be explicitly examined.

\section{Strategies for Modeling Strong Lenses}

Given a general lens solver, an important application is fitting
models to observed lenses. Lens observations now offer a wide range
of high-precision data, and this section presents a variety of
techniques for using the data to constrain models. For lenses with
point-like images, the positions, fluxes, and time delays of the
images can be used with least-squares fitting methods, and Sections
4.1--4.3 give definitions of the appropriate goodness of fit
statistics. It may be possible to use the image positions to solve
for certain model parameters analytically, as shown in Section 4.4.
Finally, for lenses with extended images like jets, arcs, and
Einstein rings, Section 4.5 briefly reviews techniques for using
them as constraints.

\subsection{Image positions}

With high-resolution optical or radio observations, it is usually
reasonable to assume that the astrometric uncertainties for
different point images are independent and Gaussian. The $\chi^2$
term for the image positions, evaluated in the {\it image plane\/},
is then
\begin{eqnarray}
  \chi^2_{img} &=& \sum_i \delta\xx_i^T \cdot S_i^{-1} \cdot \delta\xx_i\,,
    \label{eq:chiimg} \\
  \delta\xx_i &=& \xx_{obs,i}-\xx_{mod,i} \,,
\end{eqnarray}
where the sum extends over all images, and $\xx_{obs,i}$ and
$\xx_{mod,i}$ are the observed and modeled positions of image $i$.
The astrometric uncertainties for image $i$ are described by the
covariance matrix
\begin{eqnarray}
  S_i &=& R_i^T \left[\begin{array}{cc}
      \sigma_{1,i}^2 & 0 \\
      0 & \sigma_{2,i}^2 \\
    \end{array}\right] R_i\,, \\
  R_i &=& \left[\begin{array}{rr}
      -\sin\th_{\sigma,i} &  \cos\th_{\sigma,i} \\
      -\cos\th_{\sigma,i} & -\sin\th_{\sigma,i} \\
    \end{array}\right] ,
\end{eqnarray}
where the error ellipse has semi-major axis $\sigma_{1,i}$,
semi-minor axis $\sigma_{2,i}$, and position angle $\th_{\sigma,i}$
(measured East of North).

There is an alternate position $\chi^2$ that is evaluated in the
{\it source plane\/} (e.g., Kayser et al.\ 1990; Kochanek 1991a),
\begin{eqnarray}
  \chi^2_{src} &=& \sum_i \delta\uu_i^T \cdot \mu_i^T
    \cdot S_i^{-1} \cdot \mu_i \cdot \delta\uu_i \,,
    \label{eq:chisrc} \\
  \delta\uu_i &=& \uu_{obs,i}-\uu_{mod} \,,
\end{eqnarray}
where $\uu_{obs,i} = \xx_{obs_i} - \del\phi(\xx_{obs_i})$ is the
source corresponding observed image $i$, $\uu_{mod}$ is the model
source position, $\mu_i$ is the magnification tensor for image
$i$, and the covariance matrix $S_i$ is as above. The factors of
the magnification tensor are inserted because if the source plane
deviation $\delta\uu_i$ is small enough that the magnification is
nearly constant, then $\mu_i \cdot \delta\uu_i \approx \delta\xx_i$
yields an approximate image plane deviation. In other words,
$\chi^2_{src}$ is an approximate version of $\chi^2_{img}$. The
approximation can be useful because it not only avoids the
need to solve the lens equation but also allows the best model
source position to be found analytically:
\begin{eqnarray}
  \uu_{mod} &=& A^{-1} \cdot \bb\,, \\
  \mbox{where}\qquad
  A &=& \sum_i \mu_i^T \cdot S_i^{-1} \cdot \mu_i\,, \\
  \bb &=& \sum_i \mu_i^T \cdot S_i^{-1} \cdot \mu_i \cdot \uu_{obs,i}\,,
\end{eqnarray}
which is straightforward to evaluate because the matrices are
$2\times2$. However, the approximation inherent in $\chi^2_{src}$
is somewhat undesirable for two reasons. First, strictly speaking
$\chi^2_{src}$ is an accurate representation of $\chi^2_{img}$ only
if $\delta\uu$ is small; in other words, $\chi^2_{src}$ should
properly be used only if a good model is already known, not in an
initial search for a good model. Second, $\chi^2_{src}$ is not
computed with directly observable quantities. Despite these
limitations, $\chi^2_{src}$ is qualitatively correct in the sense
that for poor models it gives a large value of $\chi^2$ and thus
indicates a poor fit. In summary, the fact that $\chi^2_{src}$
is computationally fast and at least qualitatively accurate means
that it can be useful for initial modeling to find the appropriate
region of parameter space to explore. The problems with the
approximation, however, mean that it is preferable to use
$\chi^2_{img}$ for refining models to find the best-fit model and
the range of models consistent with the data.

\subsection{Image fluxes}

The (relative) fluxes of point images can provide useful model
constraints, but they must be used carefully. The problem is
that the statistical errorbars may underestimate the actual
uncertainties.  The fluxes, especially at optical wavelengths,
may be affected by microlensing, small-scale structure in the
lens galaxy, and differential reddening by dust in the lens
galaxy (e.g., Mao \& Schneider 1998; Falco et al.\ 1999;
Wozniak et al.\ 2000). They can also be affected by source
variability and time delays. If these systematic effects are not
understood and corrected, the strength of the flux constraints
must not be overemphasized. A simple and common way of using the
fluxes conservatively is to retain the assumption of Gaussian
errors for simplicity, but to inflate the errorbars to represent
an estimate of the systematic uncertainties (e.g., Koopmans \&
Fassnacht 1999; Cohn et al.\ 2000).

If the photometric errors for the various images are independent
and Gaussian, the $\chi^2$ for the fluxes is
\begin{equation} \label{eq:fluxchi}
  \chi_{flux}^2 = \sum_i { \left( f_i - \m_i f_{src} \right)^2
    \over \sigma_{f,i}^2 }\ ,
\end{equation}
where the observed flux of image $i$ is $f_i \pm \sigma_{f,i}$,
and the model gives the magnification $\m_i = |\det(\mu_i)|$ of
image $i$ and the intrinsic flux $f_{src}$ of the source.  The
best-fit source flux can be found analytically,
\begin{equation} \label{eq:fluxbest}
  f_{src} = { \sum_i f_i\,\m_i/\sigma_{f,i}^2 \over
    \sum_i \m_i^2/\sigma_{f,i}^2 }\ .
\end{equation}
If it is preferable to express the photometry in magnitudes
(rather than physical fluxes), the $\chi^2$ term is
\begin{equation} \label{eq:magchi}
  \chi_{mag}^2 = \sum_i {(m_i+2.5\log\m_i-m_{src})^2 \over \sigma_{m,i}^2 }\ ,
\end{equation}
where the observed magnitude of image $i$ is $m_i \pm \sigma_{m,i}$.
The best-fit source magnitude is then
\begin{equation} \label{eq:magbest}
  m_{src} = { \sum_i (m_i+2.5\log\m_i)/\sigma_{m,i}^2
    \over \sum_i 1/\sigma_{m,i}^2 }\ .
\end{equation}
In all of these equations the units are arbitrary; they can be
absolute fluxes or magnitudes, or they can be relative fluxes
or magnitudes referenced to one of the images.

The assumption that the uncertainties in the (relative)
photometry are independent holds only if the images do not
have overlapping point spread functions. For lenses that are
poorly resolved, deconvolution may be able to separate the
various images (e.g., Courbin et al.\ 1997), but may (in
principle) yield correlated photometric errors. The flux
$\chi^2$ is easily generalized to handle correlated errors,
\begin{equation}
  \chi_{flux}^2 = \sum_{ij} (S_f^{-1})_{ij}
    (f_i-\m_i f_{src}) (f_j-\m_j f_{src})\,,
\end{equation}
where $S_f$ is the covariance matrix of the fluxes, and both the
$i$ and $j$ sums extend over all images. The best-fit source flux
is then
\begin{equation}
  f_{src} = { \sum_{ij} (S_f^{-1})_{ij} \m_i f_j 
   \over \sum_{ij} (S_f)^{-1}_{ij} \m_i \m_j }\ .
\end{equation}
If it is preferable to use magnitudes, the $\chi^2$ term is
\begin{equation}
  \chi_{mag}^2 = \sum_{ij} (S_m^{-1})_{ij}
    (m_i+2.5\log\m_i-m_{src}) (m_j+2.5\log\m_j-m_{src})\,,
\end{equation}
where $S_m$ is the covariance matrix of the magnitudes. In this
case the best-fit source magnitude is
\begin{equation}
  m_{src} = { \sum_{ij} (S_m^{-1})_{ij} (m_j+2.5\log\m_j)
    \over \sum_{ij} (S_m^{-1})_{ij} }\ .
\end{equation}

\subsection{Time delays}

By measuring light curves for the images and comparing them to
each other, it may be possible to measure the time delay(s)
induced by lensing. In a lens with $N$ images there are up to
$N-1$ independent time delays. They can be used with lens models
to determine the Hubble constant $H_0$ (e.g., Refsdal 1964),
and to provide up to $N-2$ additional constraints on the models.
To understand how time delays are used, let $\tau_{obs,i}$ be the
time by which the light curve of image $i$ lags the light curve
of the leading image. The prediction from a lens model for the
delay can be decomposed as follows (see eq.~\ref{eq:tdel}):
\begin{eqnarray}
  \tau_{mod,i} &=& h^{-1}\,t_0\,\bar\tau_{mod,i}\,, \label{eq:tfactor} \\
  t_0 &=& {1+z_l \over c}\,{D_{ol} D_{os} \over D_{ls}}\ ,
    \qquad\mbox{(computed for $h=1$)} \\
  \bar\tau_{mod,i} &=& {1\over2}\Bigl[ \left|\xx_i-\uu\right|^2
    - \left|\xx_{lead}-\uu\right|^2 \Bigr]
    - \Bigl[ \phi(\xx_i) - \phi(\xx_{lead}) \Bigr] ,
\end{eqnarray}
where $h = H_0 / (100\ \Hunits)$ and the subscript ``lead''
indicates the leading image. \refEq{tfactor} shows that the model
time delay includes a factor that depends only on cosmology ($t_0$)
and a factor that depends only on the lens model ($\bar\tau_{mod,i}$);
it also shows the explicit dependence on the Hubble parameter $h$.

If the errors in the time delays are independent and Gaussian, the
$\chi^2$ contribution is
\begin{equation} \label{eq:chitdel}
  \chi^2_{tdel} = \sum_i { \left( \tau_{obs,i} - h^{-1}\,t_0\,\bar\tau_{mod,i}
    \right)^2 \over \sigma_{t,i}^2 }\ .
\end{equation}
To find the best-fit value of $h$ together with confidence
intervals, one approach is to trace out $\chi^2$ versus $h$.  An
alternative approach is to let $h$ be free to adopt the best-fit
value, although perhaps with a prior range
$h_{prior}\pm\sigma_{h,prior}$ specified to ensure that the value
is reasonable. (To let $h$ be unrestricted, i.e.\ with no prior
range specified, simply use a large value for $\sigma_{h,prior}$.)
The $\chi^2$ contribution for the prior range is
\begin{equation}
  \chi^2_h = { \left( h^{-1} - h^{-1}_{prior} \right)^2 \over
    \sigma_h^2 }\ ,
  \qquad\mbox{where}\qquad
  \sigma_h = {\sigma_{h,prior} \over h_{prior}^2}\ .
  \label{eq:chih}
\end{equation}
Note that this $\chi^2$ contribution is written in terms of
$h^{-1}$ rather than $h$, and $\sigma_h$ is the uncertainty in
$h^{-1}$, while $\sigma_{h,prior}$ is the uncertainty in $h$
itself. This is done for convenience, making it possible to solve
analytically for the best-fit value of $h$ (the value that
minimizes $\chi^2_{tdel} + \chi^2_h$),
\begin{equation}
  h =
    \left[ {1 \over \sigma_h^2} +
    \sum_i {t_0^2\,\bar\tau_{mod,i}^2 \over \sigma_{t,i}^2} \right]
    \Bigg /
    \left[ {1 \over h_{prior}\,\sigma_h^2} +
    \sum_i {t_0\,\bar\tau_{mod,i}\,\tau_{obs,i} \over \sigma_{t,i}^2} \right] .
    \label{eq:besth}
\end{equation}
With a single measured time delay $\tau_{obs}$ and no prior
assumption about $H_0$, \refeq{besth} is equivalent to $h =
t_0\,\bar\tau_{mod}/\tau_{obs}$. If there are $M$ time delays,
they determine a value for $h$ and also provide $M-1$ constraints
on lens models.

A potential problem with this analysis is that the time delay
errors may be very non-Gaussian. The light curves used to measure
time delays usually have irregular sampling and may have
substantial non-Gaussian ``noise'' from phenomena such as
microlensing (e.g., Thomson \& Schild 1997; Fassnacht et al.\
1999). A relatively simple test can be done to examine how
conclusions about the Hubble constant are affected by non-Gaussian
errors. One way to estimate the error distribution for the time
delays is to perform Monte Carlo simulations with the light
curve data; the simulations yield a set of time delays whose
distribution represents the uncertainties. Using all of the
simulated time delays to determine the Hubble constant produces
a distribution of $H_0$ values that incorporates all of the
time delay errors, including not only non-Gaussian properties
but also any correlations that might exist. Fassnacht et al.\
(1999) and Koopmans \& Fassnacht (1999) have performed this test
using the three time delays measured for the 4-image lens
B~1608+656.  Assuming independent non-Gaussian errors, i.e.\
using \refeq{chitdel}, they find $H_0 = 59_{-6}^{+7}\ \Hunits$;
by contrast, when they use the full error distribution from
Monte Carlo simulations they find $H_0 = 59_{-7}^{+8}\ \Hunits$.
(Both sets of errorbars encompass the 95\% confidence range.)
While this single test cannot rule out the importance of
non-Gaussian time delay errors in all lensing determinations of
$H_0$, it does suggest that their effects may not be dramatic.

\subsection{Linear parameters and constraints}

In parametric lens models, the lensing potential, deflection, and
magnification are often complicated functions of the parameters.
Sometimes, however, there are parameters that enter as simple
linear coefficients. For example, the potential of a softened
isothermal sphere lens is
\begin{equation} \label{eq:phiiso}
  \phi(r) = b \left[ \sqrt{s^2+r^2} - s -
    s\,\ln\left({s+\sqrt{s^2+r^2} \over 2s}\right) \right] .
\end{equation}
The core radius $s$ is a non-linear parameter because it appears in
the arguments of transcendental functions. The mass parameter $b$,
by contrast, is a linear parameter because it appears as a simple
multiplicative coefficient. For non-linear parameters, the only way
to find the best-fit values is to use a numerical algorithm to find
minima of the multi-dimensional $\chi^2$ function. For linear
parameters, however, it may be possible to compute their best-fit
values analytically, thus reducing the number of parameters that
must be varied in the numerical optimization.

In principle, position, flux, and time delay constraints can all be
used to solve for linear parameters. In practice, however, the flux
and time delay data are usually too unreliable to be used this way.
Besides, the image positions often provide more than enough
constraints to solve for all of the linear parameters in a model,
so this section discusses only position data. Linear techniques
have been used by, e.g., Kochanek (1991b), Bernstein \& Fischer
(1999), and Keeton et al.\ (2000a), but this section offers a more
general discussion.

Consider $N$ images $\{\xx_i\}_{i=1}^{N}$ of a given source.
Requiring that the images be fit {\it exactly\/} gives $N$ vector
equations,
\begin{equation}
  \uu = \xx_i - \del\phi\left(\xx_i\right)\,, \qquad (i=1,\ldots,N)\,.
\end{equation}
Eliminating the unobservable source position, this is equivalent to
$N-1$ vector equations
\begin{equation}
  \xx_1 - \del\phi\left(\xx_1\right) = \xx_j - \del\phi\left(\xx_j\right)\,,
  \qquad (j=2,\ldots,N)\,. \label{eq:lincon1}
\end{equation}
Because the vector space is two-dimensional, this amounts to
$2(N-1)$ constraint equations.

Now suppose that some of the parameters in the lens model are
linear, i.e.\ they enter as simple multiplicative coefficients in
the lensing potential (like $b$ in eq.~\ref{eq:phiiso}). Let the
$M$ linear parameters be $\pp = \{p_\al\}_{\al=1}^{M}$ and the $M'$
non-linear parameters be $\qq = \{q_\bt\}_{\bt=1}^{M'}$. The
definition of linear parameters is that the lensing potential can
be decomposed as
\begin{equation}
  \phi(\xx;\pp,\qq) = \sum_{\al=1}^{M} p_\al\,\varphi_\al(\xx;\qq)
  \label{eq:linparm1}
\end{equation}
for some set of functions $\{\varphi_\al(\xx;\qq)\}_{\al=1}^{M}$
that are functions of position and are parametrized by only the
non-linear parameters $\qq$.

If the potential has the form \refeq{linparm1}, then the
constraints in \refeq{lincon1} become
\begin{equation}
  \xx_1 - \sum_{\al=1}^{M} p_\al\,\del \varphi_\al(\xx_1;\qq) =
  \xx_j - \sum_{\al=1}^{M} p_\al\,\del \varphi_\al(\xx_j;\qq)\,,
  \qquad (j=2,\ldots,N)\,. \label{eq:lincon2}
\end{equation}
If the number of constraints equals the number of linear
parameters, i.e.\ $M=2(N-1)$, then \refeq{lincon2} is a square
matrix equation that can be solved to find the set of linear
parameters $\pp$ that yield an exact fit to the image positions.
Because the constraint equations depend on the non-linear
parameters $\qq$, solving \refeq{lincon2} actually gives the linear
parameters as functions of the non-linear parameters, $\pp(\qq)$.
Formally, this amounts to solving the matrix equation
\begin{equation}
  A(\qq) \cdot \pp(\qq) = \bb\,, \label{eq:lincon3}
\end{equation}
where the matrix $A$ has the components $A_{i\al}$ and the vector
$\bb$ has the components $b_i$ such that
\begin{eqnarray}
  A_{(2j-3)(\al)} &=& {\partial \over \partial x} \varphi_\al(\xx_j;\qq)
    - {\partial \over \partial x} \varphi_\al(\xx_1;\qq)\,,
    \qquad (j=2,\ldots,N) \\
  A_{(2j-2)(\al)} &=& {\partial \over \partial y} \varphi_\al(\xx_j;\qq)
    - {\partial \over \partial y} \varphi_\al(\xx_1;\qq)\,, \\
  b_{(2j-3)} &=& x_j - x_1\,, \\
  b_{(2j-2)} &=& y_j - y_1\,.
\end{eqnarray}

We can generalize this technique to sets of images associated with
different sources. Suppose there are $N$ images
$\{\xx_i\}_{i=1}^{N}$ of one source, and $N'$ images
$\{\xpp_i\}_{i=1}^{N'}$ of a second source. Then the full set of
constraints is
\begin{eqnarray}
  \xx_1  - \del\phi\left(\xx_1 \right) &=& \xx_j  - \del\phi\left(\xx_j \right)\,,
    \qquad (j=2,\ldots,N)\,, \\
  \xpp_1 - \del\phi\left(\xpp_1\right) &=& \xpp_k - \del\phi\left(\xpp_k\right)\,,
    \qquad (k=2,\ldots,N')\,,
\end{eqnarray}
(compare eq.~\ref{eq:lincon1}). Again finding the proper number of
linear parameters, now $M=2(N+N'-2)$, leads to a solvable matrix
equation similar to \refeq{lincon3}, where the generalization of
the matrix $A$ and vector $\bb$ is straightforward: simply insert
the entries for the second set of images $\{\xpp_i\}_{i=1}^{N'}$ as
additional rows in $A$ and $\bb$. The generalization to further
sets of images is similar.

When used properly, linear techniques can speed up model searches
by eliminating some parameters and by automatically identifying
models that match at least some of the data. These features are
especially helpful when working with high-precision data, because
good models are likely to be confined to a narrow range of parameter
space, so the $\chi^2$ surface is likely to be a high plateau cut
by narrow valleys in a way that frustrates multi-dimensional
optimization algorithms (see Press et al.\ 1992). Without linear
techniques, the optimization routine may wander around and miss
the valleys; but with linear techniques, the optimization routine
is directly guided to the valleys that need to be explored. Two
important points must guide the use of linear techniques, however.
The first is that number of linear parameters and constraints must
match. For example, the relative positions in a 2-image lens
provide two constraints, so they require two linear model
parameters to be used with linear techniques. The relative
positions in a 4-image lens provide six constraints, so they
require six linear parameters.

The second important point is that using linear techniques forces
the model to fit the data {\it exactly\/}, which is much stronger
than asking the model to reproduce the data within the errorbars.
Consider a lens (call it $A$) produced by a smooth mass
distribution. Also consider a second lens (call it $A'$) where the
same smooth mass distribution is perturbed by small-scale structure
such as an isophote twist or a lumpy mass distribution. The
small-scale structure can cause the image positions in lens $A'$ to
differ from those in lens $A$ by up to $\sim$1 milli-arcsecond,
even for the same source position (see Mao \& Schneider 1998).
Suppose the observational uncertainties are $\gtrsim$1 mas, so a
simple smooth lens model yields a perfect fit to lens $A$ and a
worse but statistically acceptable fit to lens $A'$. Now if linear
techniques are applied to lens $A'$, the smooth model will be
forced to adjust large-scale mass components to try to fit the data
exactly. The model may be pushed into a strange corner of parameter
space, or it may simply report that it cannot fit the data. Either
way, the modeler will be led to believe that there are significant
differences between the lens galaxies in $A$ and $A'$, when in fact
the only difference is in small-scale mass components. A similar
argument holds even when linear techniques are not used, if the
observational uncertainties are much smaller than 1 mas.

The mistake here is the use of an oversimplified lens model with
high-precision data. The lesson is that because of the possibility
of perturbations from substructure, simple smooth lens models should
not be asked to fit data to better than $\sim$1 mas precision. In
other words, when using high-precision data in general, and
specifically when using linear techniques, it is very important to
allow complexity in the lens model. As an example, Keeton et al.\
(2000a; also see Kochanek 1991b; Bernstein \& Fischer 1999) show
that allowing physically-motivated substructure in the models is
essential for robustly measuring the large-scale properties of the
lens galaxy in the lens Q~0957+561.

\subsection{Extended images}

Extended images like arcs or Einstein rings can dramatically
increase the number of constraints on models (compared with the
constraints from point images). Extended images must be treated
with some care, though, and several techniques have been developed
to deal with them.  Some radio images that appear point-like in
low- or moderate-resolution radio maps can be resolved into
sub-components by high-resolution VLBI or VLBA maps. The
sub-components can be used as multiple sets of point-like images
that have high-precision astrometry. See Trotter et al.\ (2000)
for an excellent recent example of VLBI data and modeling. For
smooth extended images like arcs and Einstein rings, several
``exhaustive'' algorithms have been developed to analyze the full
observed map while allowing for the fact that the intrinsic source
is unknown (e.g., Kochanek et al.\ 1989; Kochanek \& Narayan 1992;
Wallington, Kochanek \& Narayan 1996; Tyson et al.\ 1998; Keeton
et al.\ 2000a). These techniques extract the maximum amount of
information from the images, but they are computationally expensive.
Kochanek et al.\ (2001ab) have developed two alternate techniques
--- ``ring fitting'' for optical or infrared Einstein rings, and
``curve fitting'' for arcs --- that extract less information from
the images but are computationally fast. At least one of these
techniques for extended images should be appropriate for any
particular lens, and together with the various techniques for
handling point-like images they make it possible to use virtually
any kind of observational data to constrain lens models.

\section{A Software Package}

The techniques presented in this paper have been used to create a
versatile software package called \gravlens\ that can be used for a
wide variety of lensing calculations. The code has a command-driven
interface that is easy to use, and it comes with a detailed manual.
The software and documentation can be downloaded from the web site
for the CfA/Arizona Space Telescope Lens Survey, at {\tt
http://cfa-www.harvard.edu/castles}.

The software is actually a package of two applications. One
application, called \gravlens, provides capabilities for basic
lensing calculations. It combines the tiling algorithm from \S 3
with an extensive catalog of circular and elliptical mass models
that can be arranged into arbitrarily complex composite models.
Thus, it can be used to study the general lensing properties of
virtually any mass distribution. It has been used by Keeton \&
Kochanek (1998), Keeton, Mao \& Witt (2000b), and Rusin et al.\
(2000) to analyze the caustic structures in complex lens models.

The second application, called \lensmodel, begins with all of the
capabilities of \gravlens\ and adds many features for modeling
observed lenses (as summarized in \reftab{lensmodel}). The
\lensmodel\ software allows any or all of the observational
constraints discussed in \S 4, so it can be used with complex
lenses that include point images of multiple sources and/or
extended images like arcs or rings. It allows arbitrarily
complicated mass models to be fit to the data, so it can be used
for essentially any lens in any environment. It gives the user
complete control over the behavior of the model parameters; for
example, certain parameters can be varied while others are held
fixed, constraints can be placed on parameter ranges, and explicit
relations can even be imposed between the parameters. The parameter
controls allow a thorough understanding of any covariances,
degeneracies, or other systematic uncertainties in the models
and the conclusions drawn from them. In other words, \lensmodel\
offers the ability to fit complex mass models to sophisticated
data, and it provides the control needed to fully analyze the
range of models consistent with the data.  The \lensmodel\
software has been used to study models of many observed lenses
(Keeton \& Kochanek 1997; Keeton et al.\ 1998, 2000a; Cohn et
al.\ 2000; Kochanek et al.\ 2000b; Leh\'ar et al.\ 2000; Ros et
al.\ 2000; and Rusin et al.\ 2000).

\acknowledgements
Acknowledgements: The techniques and software presented here have
been under development for several years, and during that time many
people have contributed to the work. Chris Kochanek has offered a
tremendous number of ideas, and has thoroughly tested each feature
as it was added to the code. Joanne Cohn, Jose Mu\~noz, David
Rusin, Brian McLeod, and Joseph Leh\'ar have repeatedly put the
code through its paces, uncovering bugs of many shapes and sizes
and offering helpful suggestions for the code and documentation.
Finally, an anonymous referee gave an insightful critique that
improved the presentation of this paper. Support for this work has
been provided by ONR-NDSEG grant N00014-93-I-0774, NSF grant
AST-9407122, NASA ATP grant NAG5-4062, and Steward Observatory.



\clearpage

\input tab_lensmod.tex


\clearpage

\end{document}

%% file: tab_lensmod.tex
\begin{deluxetable}{ll}
\tablewidth{0pt}
\tablecaption{Features of \lensmodel}
\tablehead{
  \colhead{Feature} & \colhead{Capability}
}
\startdata
Point images
 & Position, flux, and time delay data. \\
 & Image plane or source plane $\chi^2$. \\
 & Can handle multiple sources. \\
\\
Arcs
 & Curve fitting. \\
\\
Rings
 & Ring fitting with elliptical sources. \\
\\
General data
 & Allows a floating registration between different \\
 & types of data (e.g., radio and optical). \\
\\
Mass models
 & Includes numerous circular and elliptical models, \\
 & and allows arbitrary combinations of them. \\
\\
Parameters
 & Gives full control over which parameters vary. \\
 & Can produce arbitrary parameter surveys. \\
 & Allows external constraints on parameter ranges. \\
 & Allows relations between parameters to be imposed. \\
 & Can use linear parameters and constraints. \\
\\
Cosmology
 & Allows arbitrary values of $\Omega_M$, $\Omega_\Lambda$, and $H_0$. \\
 & Can determine $H_0$ from time delays. \\
\\
Output
 & Goodness of fit. \\
 & Properties of mass model. \\
 & Critical curves and caustics. \\
 & Plots of time delay surface, potential, etc. \\
\enddata
\label{tab:lensmodel}
\end{deluxetable}